# Effect of Variation of Different Nano filler on Structural, Electrical, Dielectric and Transport Properties of Blend Polymer Nanocomposites

Anil Arya[1], Mohd Sadiq[2,1], A. L. Sharma[1*]

[1]Centre for Physical Sciences, Central University of Punjab, Bathinda-151001, Punjab, INDIA

[2]Department of Physics, A R S D College, Delhi -11021, India

Corresponding Author E-mail: alsharmaiitkgp@gmail.com

**Abstract**

In the present work, the effect of different nanofiller ($BaTiO_3$, $CeO_2$, $Er_2O_3$ or $TiO_2$) on blend solid polymer electrolyte comprising of PEO and PVC complexed with bulky $LiPF_6$ have been explored. The XRD analysis confirms the polymer nanocomposite formation. FTIR provides evidence of interaction among the functional groups of the polymer with the ions and the nanofiller in terms of shifting and changing peak profile. The highest ionic conductivity is $\sim 2.3 \times 10^{-5}$ S cm$^{-1}$ with a wide electrochemical stability window of ~3.5 V for 10 wt. % $Er_2O_3$. The real and imaginary part of dielectric permittivity follows the identical trend of the decreasing value with increase in the frequency. The particle size and the dielectric constant shows an abnormal trend with different nanofiller. The ac conductivity follows the universal power law. An effective mechanism has been proposed to understand the nanofiller interaction with cation coordinated polymer in the investigated system.

**Introduction**

The demand of miniaturized power supply with desirable properties in electrochemical devices has been rapidly increased since last two decades. Out of a large number of electrochemical energy storage devices, rechargeable lithium ion batteries are most intensively studied for portable electronic gadgets. The frequent use of those devices gears the researchers toward the remarkable development of secondary rechargeable battery with high capacity, light weight and lower cost in contrast to primary batteries. As, the electrolyte is an essential component in lithium ion battery as it separates both electrodes (anode and cathode), controls the number of charge carriers and permits their movement during charging/discharging process. Ionically conducting polymer system with polymer as separator/electrolyte are of intense interest due to their multiple applications, including super capacitors, fuel cell, solar cells, electro chromic windows, and Li ion batteries [1-4] The desirable electrolyte for any application in energy storage/conversion devices must have (a) high ionic conductivity; (b) electrochemical stability window (>4 V); (c) low melting point; (d) high boiling point; (e) high chemical stability; (f) nontoxicity; (g) low cost and good compatibility with electrodes. Solid polymer electrolyte (SPEs) plays a dual role as an electrolyte as well as a separator which keeps both electrodes separate and avoids the user from using a spacer as in liquid electrolyte system.

The first report on ionic conduction was given by P. V. Wright [5] and Fenton et al [6] in 1973 and it was concluded that polymer host dissolved with an alkali metal salt results in an ionic conductive system. The first application of novel polymer electrolyte in batteries was announced by Armand and his co-workers in 1978 [7]. Most devices are based on liquid/gel polymer electrolytes due to their high ionic conductivity ($10^{-3}$-$10^{-2}$ Scm$^{-1}$) and compatible with electrodes. But poor mechanical strength, freezing at low temperature, leakage and flammable nature limits their application in commercial use. This motivates the researchers toward better replacement of liquid polymer electrolyte with a solvent free polymer electrolyte having high ionic conductivity, leak proof, flexibility, wide electrochemical window, good mechanical strength, light weight and ease of preparation [8-10]. Solid polymer electrolytes (SPEs) fulfill all above requirements and have attracted researchers globally towards development for their potential application in portable electronics and electrochemical devices [11].

The commonly used host polymer for preparation of polymer nanocomposite films (PNCs) is PEO [12], PAN [13], PMMA [14], PEMA [15] and PVC [16]. Out of aforesaid host polymers, polyethylene oxide (PEO) is undoubtedly the best host polymer used as SPEs with strong, unstrained C-O, C-C, C-H bonds and it has high dielectric constant, easy availability, and high ionic conductivity in amorphous phase, low glass transition temperature, high flexibility, and good dimensional and chemical stability. But PEO possesses low ionic conductivity due to the presence of crystalline domains when complexed with alkalis salts. There are various techniques for improving the amorphous phase, electrical and mechanical properties of SPEs like crosslinking/blending of two or more polymers and the addition of nanofiller/plasticizer in the polymer electrolyte. The blending of one polymer with another is a promising approach as it provides in hand the unique and superior tunable properties for application than single polymer [17-18]. Also, the easy control and modulation of properties can be achieved by varying the composition of the individual polymer [19-20].

PEO has high molecular weight and ionic conductivity of the order of $10^{-7}$ S cm$^{-1}$ because of the lone pair of electrons in its backbone with a lithium salt. The low conductivity of pure PEO is due to semi crystalline nature at ambient temperature which restricts the ion motion. But above the melting temperature crystalline to amorphous region transition occurs which enhances the conductivity. The movement of charge carriers (cation/anion) occurs in amorphous phase supported by the segmental motion of polymer host. Polyvinyl chloride (PVC) is chosen as a secondary polymer, it acts as a mechanical stiffener in the electrolyte due to its immiscibility. The use of PVC as the blend is attributed mainly to the presence of lone pair electrons at the chlorine atom where the inorganic salts can be solvated. The addition of salt in polymer matrix enhances the conductivity via interaction with polymer host. The low lattice energy salt with small polarizing cation and the large anion of delocalized charge is chosen in order to meet the desirable requirement. $LiPF_6$ is a better choice among the tried alkali salt due to larger anion size (~2.8 Å), weak $Li^+$---$PF_6^-$ interactions, low ion pair formation ability, good solubility, highly electropositivety, flame retardant, low cost and good environmental impact [21, 22]. The larger anion size supports the faster ion dissociation via strong columbic interaction forces between oppositely charged ions. The addition of nanofiller in SPEs is the preeminent approach to improve the electrical, thermal and mechanical properties as reported by many researchers [12, 23-24]. Generally nanofiller reduce the crystallinity of polymers and amorphous phase formation increases the ionic conductivity by penetrating into polymer matrix [25]. The addition of filler destructs the dense packing and crosslinking density of polymer by retarding the recrystallization affinity of polymer which provides the favorable conduction path for ion migration on the surface of nanofiller. The presence of ceramic fillers in PNCs effects the polymer dipole orientation, which provides flexible polymer segmental chain motion as well as dissociation of ion pairs due to the surface charge on the ceramic filler [26].

Incorporation of the nanofiller helps in inhibiting the recrystallization of the polymer and decrease the glass transition temperature of the composite polymer electrolyte and hence enhanced conductivity is achieved [27]. The enhancement of ionic conductivity is attributed to Lewis acid-base type interactions of conductive species with the surface group of the nanofiller which decrease the polymer reorganization tendency and modifies the polymer chain arrangement. In addition, nanofiller also supports ion dissociation and enhance the ion migration by providing additional conducting pathways within host polymer matrix [28-29]. Figure 1 shows the molecular structures of the materials. So, in regards to enhancement in properties the particle size and dielectric constant of the nanofiller plays the crucial role.

The present paper report is mainly concerned with the effect of various nanofiller having a different particle size and the dielectric constant on the polymer blend electrolyte. To obtain the insights in the above said parameters different characterization techniques such as; X-Ray diffraction, FTIR, FESEM, Impedance study are performed. The impedance data is transformed into the dielectric data to understand the effect of the dielectric constant of nanofiller in enhancing the overall dielectric constant of the whole polymer matrix. Also, the ionic transference number and the electrochemical stability window was obtained using i-t characterization technique and the linear sweep voltammetry respectively.

**Experimental**

*Materials*

The SPEs system involved the use of poly (ethylene oxide) (PEO) and poly (vinyl chloride) (PVC) with average molecular weights of $1 \times 10^6$ (Sigma Aldrich), and $6 \times 10^4$ respectively. $LiPF_6$ was purchased from Sigma Aldrich. Titanium oxide ($TiO_2$), Barium Titanate ($BaTiO_3$), Cerium oxide ($CeO_2$) and Erbium oxide ($Er_2O_3$) were acquired from Sigma Aldrich. Anhydrous tetrahydrofuran (THF) was used as solvent purchased from Sigma Aldrich. Table 1 shows the particle size and crystalline size for the investigated nanofiller in the present system.

*Preparation of solid polymer electrolyte*

All the solid polymer electrolyte (SPEs) samples were prepared using universal solution cast technique. A snapshot of the technique is shown in the flowchart (Fig 2). The PEO is used as host polymer and PVC as blend polymer in constant ratio 50:50. $LiPF_6$ was chosen as salt in constant ratio [Ö]:[Li$^+$] of 8:1 (eq. 1) and various nanofiller $BaTiO_3$, $TiO_2$, $CeO_2$ and $Er_2O_3$ with 10 wt. % constant concentration w.r.t host polymer PEO.

$$\frac{\text{Ö}}{\text{Li}^+} = \frac{\text{No. of monomer unit in half a gram of PEO}}{\text{No. of } LiPF_6 \text{ molecular in half a gram of salt}} \times \frac{\text{wt. of PEO was taken}}{\text{wt. of salt taken}} \quad (1)$$

The preparation of SPEs involves the dissolution of PEO, PVC in tetrahydrofuran (THF) solvent at 40 °C and then polymer solutions were mixed/stirred for several hours to form the polymer matrix of electrolyte system. After complete dissolution of the polymer, an appropriate amount of $LiPF_6$ was added to the polymer blend matrix and the mixture was stirred continually for good polymer salt complex system for 12 hr. Later in polymer salt complex various nanofiller is added in polymer salt system and were stirred up to 8 hr. in order to obtain a homogeneous solution. The final solution was sonicated for 30 minutes, then poured in glass petri dishes and left for drying by solvent evaporation at room temperature in a vacuum oven. The resultant films formed were kept for further characterization. The samples are designated as PP (pure polymer blend), PPL (pure polymer with $LiPF_6$ salt), PPLB (with $BaTiO_3$ nanofiller), PPLC

(with $CeO_2$ nanofiller), PPLE (with $Er_2O_3$ nanofiller), PPLT (with $TiO_2$ nanofiller). Some most intrinsic properties of various procured nanofiller are also tabulated in table 1.

*Characterization*

**X-ray diffraction (XRD) analyses**

X-ray diffraction (XRD) is the most useful method for the determination of crystal structure and crystallinity of the materials. The X-ray diffraction studies were performed with 2θ ranging from 10° to 50°. The X-ray source is Cu-K$_α$ radiation with wavelength λ=1.54 Å.

**Field Emission scanning electron microscope (FESEM) analysis**

Field emission scanning electron microscopy (FESEM) was used to characterize the morphology of the solid polymer electrolytes using Carl Zeiss product. Films were gold coated and examined under high vacuum.

**Fourier transforms infrared (FTIR) spectroscopy**

FTIR spectroscopy was carried out on all solid polymer electrolyte films using the Bruker Tensor 27 (Model: NEXUS –870)) in absorbance mode over the wavenumber region from 600 to 4000 cm$^{-1}$ with a resolution of 4 cm$^{-1}$.

**Electrochemical impedance spectroscopy (EIS)**

Impedance measurements of the solid polymer electrolyte films were carried out using CHI 760 Electrochemical Analyzer over a frequency range between 1 Hz to1 MHz at room temperature. Each sample was sandwiched between two stainless steel (SS) electrodes and an AC signal of 50 mV was applied for measurement. The intercept between the semi-circle at high frequency and tilted spike at low frequency were taken as the bulk resistance ($R_b$). The ionic conductivity (σ) value was obtained using equation 2:

$$\sigma_{dc} = \frac{1}{R_b}\frac{t}{A} \qquad (2)$$

Where t is thickness (cm) of polymer film, $R_b$ is bulk resistance (Ω) and A is area (cm$^2$) of working electrode.

**Ionic Transference number**

The stainless steel/polymer electrolyte/stainless steel (SS|PE|SS) symmetric cells were used in the transference number measurements similar to those described in the preceding paragraph. Ion transference numbers of the solid polymer electrolytes were determined by DC Polarization technique using equation 3 & 4:

$$t_{ion} = \left(\frac{I_t - I_e}{I_t}\right) \times 100 \qquad (3)$$

and

$$t_{ion} + t_e = 1 \qquad (4)$$

$I_t$ and $I_e$ are the total and residual current respectively

**Voltage Stability**

The voltage stability of solid polymer electrolyte films was obtained using the linear sweep Voltammetry technique, and cyclic voltammetry measurements were performed at a scan rate of 10 mV/s in CHI 760.

**Dielectric analysis**

The real (Z′) and imaginary (Z″) parts of the complex impedance were used to express the real (ε′), imaginary (ε″) parts of permittivity, ac conductivity and real and imaginary part of the modulus (M).

**Results and discussion**

**X-ray diffraction (XRD) analysis**

The XRD patterns of the pure polymer blend (PP) and PPL, PPLB, PPLC, PPLE, PPLT are depicted in Figure 3. XRD pattern of pure PEO-PVC film shows two sharp partially crystalline peaks at 2θ =14.01° and 16.81°. The addition of salt to the pure polymer blend shows shifting in peaks to 14.05° and 16.87°. Besides these crystalline peaks, the broad hump between 12° and 18° is also observed, and it arises from the dominant amorphous phase of the PVC [30]. Two signature peaks are observed corresponding to PEO one with low intensity at 2θ =18.34° and 23.66° which get shifted with the addition of salt to 18.70° and absence of second peak suggesting suppression of crystallinity in polymer matrix due to the strong interaction of lithium ion with polymer blend. The indexing of two peaks is reported to be [120] at 18.34° and [032]+[112] at 23.66° [31].

The $LiPF_6$ exhibits two strong peaks at 21°−26° and minor peaks at 42°, 51°, 57° [32]. The absence of these peak in the polymer salt system evidences the complete dissociation of salt in the polymer matrix. The interaction of cation with the ether group of PEO separates the cation and anion which enhances the amorphous content. The above examination shows the presence of multiphase character consisting of both crystalline and amorphous nature [33-34]. Further addition of nanofiller may alter the polymer amorphous content due to the formation of the cation-nanofiller and the polymer nanofiller interaction.

The addition of nanofiller in polymer salt matrix indicated backwards shift in PVC and PEO peaks toward lower angle side (2θ =13.88°, 16.68°) confirming the effect of nanofiller on PS complex. The addition of nanofiller has brought noticeable changes in the peaks with a shift in position and broadening. The shift of PEO peak toward lower angle side on the addition of Barium Titanate nanofiller suggests an increase in d-spacing of host PEO. This provides direct evidence of the presence of strong interaction between nanofiller and polymer salt complex. Figure 3b shows the peaks for $BaTiO_3$ at 2θ =32.08°, 33.67°, 36.47°, 39.18°, 43.44° and 46.81° indexed as [110/101], [001/100], [110], [111], [200] and [201] respectively [35-36]. The addition of $CeO_2$ nanofiller confirms the polycrystalline nature of the SPEs and high intensity peaks were observed at 33.80°, 47.90° respective to the [200], [220] crystal planes [37]. The XRD spectra of $Er_2O_3$ doped nanofiller shows the signature of poor crystallization peak of nanofiller at 46.65° indexed as [440]. Other peaks are at 32.0°, 36.81° indexed as [400] and [400] [38]. The $TiO_2$ nanofiller added SPE is shown in Fig 3 b indicated as PPLT. XRD pattern exhibits lower intensity peaks at 25.52°, 31.38°, 34.05°, 37.26°, 41.27°, 42.43°, 46.23°, 48.99° respective to [101], [121], [003], [004], [111], [221], [032] and [200] planes [39-40]. It is concluded form the above that the characteristics peak of the nanofiller are in good agreement with the previously reports and the less intense peak indicates the semi-crystalline nature of nanofiller. The d-spacing between the diffraction planes was obtained using the Bragg's formula $2d\sin\theta=n\lambda$ and interchain separation (R) using the equation $R=5\lambda/8\sin\theta$ and determined values are shown in Table 2. The average crystalline was determined by Debye Scherrer's equation $L=0.94\lambda/\beta\cos\theta$ and the $BaTiO_3$, $CeO_2$, $Er_2O_3$ and $TiO_2$ have size 12.56, 4.07, 8.40 and 5.24 nm. All the nanofiller dispersed system shows the characteristics peak of nanofiller which indicates the formation of ion-dipolar complexes due to ion-polymer-nanofiller interactions. From the different structural properties of nanofiller it is expected that their effect on the structural and electrical properties will be different. From the Table 2 it can be noticed that the d-spacing value varies slightly with the addition of different nanofiller while the interchain separation shows noticeable change and maximum for the $Er_2O_3$ nanofiller.

**Field emission scanning electron microscope (FESEM) analysis**

Surface morphology studies were carried out by field emission scanning electron microscopy (FE-SEM) to investigate the effect of nanofiller content on PEO-PVC based solid polymer electrolyte. FESEM micrographs of different nanofiller doped SPEs are represented in Fig 4. The FESEM image of pure blend polymer shows rough surface morphology with disturbed smoothness suggesting the presence of crystallinity in the polymer matrix (Fig 4 a). The absence of any sharp interface between PEO and PVC polymer suggests good compatibility between both polymers. Further addition of salt in the pure polymer matrix changes the texture and a smooth surface with small pores are obtained which is an indication of complete dissolution of salt in blend polymer. It reveals the formation of complex and is in good agreement with XRD analysis.

The surface morphology of blend polymer after addition of salt is shown in Fig 4 b. Further addition of nanofiller modifies the surface morphology and texture. The smooth morphology for the erbium oxide nanofiller based SPE associated with the enhancement of amorphous phase. The presence of white spots on the surface of the polymer films indicates the uniform and homogenous distribution of erbium oxide nanofiller in the whole polymer salt matrix (Figure 4 c). No phase separation is visible in this micrograph suggesting the rapid transport of ions without the presence of any blocking phase [41]. This type of nature suggests the presence of strong interaction between polymer, salt and nanofiller. The addition of Barium Titanate nanofiller changes shape from spherical to rectangular particles. Figure 4 (e) shows a morphology which may be attributed to the formation of $TiO_2$ rich polymer/salt aggregates. A phase separation occurs here which hinders the ion transport due to phase blockage of ions (Figure 4 c & d) and a less ion transport is expected here which leads to a lower value of conductivity as discussed in later section.

On the other hand, Fig. 5 shows the EDX elemental analysis of pure polymer and nanofiller doped solid polymer electrolytes. The Y-axis shows the counts (number of X-rays received and processed by the detector) and the X-axis shows the energy level of those counts. This result can be used as evidence to confirm the existence of nanofiller in the electrolyte membrane after the stirring, mixing and drying processes.

**Fourier Transform Infrared (FTIR) Analysis**

FTIR spectroscopy was used for investigating the microscopic details of SPEs and was expected to provide interactions such as polymer ion interaction, ion-ion interaction and polymer ion filler interaction. Figure 6 shows the FTIR spectra in absorbance mode of SPE films in the wavenumber region (600-3000 $cm^{-1}$) and band assignment of experimentally observed bands are given in Table 3. The PEO-PVC spectrum shows various absorption bands as a fingerprint in the energy region between 600 and 1500 $cm^{-1}$ (Figure 6). The characteristic absorption modes observed in the FTIR spectrum at the wavenumbers ~732 , ~842, ~950, ~1048 , ~ 1113, ~ 1244 ,~ 1351,~ 1432, ~ 1468, ~2884 and ~2939 $cm^{-1}$ are attributed to C-Clst., $\nu_s(PF_6^-)$, (C-O-C)st., $\nu(COC)_s \tau(CH_2)_s, w(CH_2)_s, \delta(CH_2)_s, \delta(CH_2)_a, \nu(CH_2)_s$ *and* $\nu(CH_2)_a$ respectively [42-43]. The peaks appearing near 720 $cm^{-1}$ corresponds to the C-Cl stretching mode and at 1048 $cm^{-1}$ assigned to (C-O-C) stretching mode of PVC [44]. The effect of the nanofiller on the peak pattern in terms of asymmetry and peak position has been recorded in the Table 3. The signature peak of host polymer PEO near 1113 $cm^{-1}$ is attributed to C-O-C symmetric stretching mode concerning ether oxygen of PEO and provides direct evidence of complexion regarding band shifting on addition of salt. So, the anion peak can be used to investigate the effect of nanofiller on the environment of the cation. Table 3 shows the shift, peak broadening in C-O-C band occurring and is strong evidence of cation coordination ($Li^+$) with ether group of PEO.. Any interaction of ($PF_6^-$) anion changes its symmetry from $O_h$ point group to $C_{3v}$ or $C_{2v}$. These features and changes in absorption mode of polymer matrix reflect their imprint in FTIR spectrum between 600 $cm^{-1}$ to 1500 $cm^{-1}$. Also, the complexation of salt with the polymer blend leads changes in polymer backbone due to coordination of cation with polymer host. Further, addition of nanofiller in polymer salt matrix also plays an active role in polymer-ion and polymer-ion-nanofiller interactions in SPEs due to surface group of nanofiller with acid base interactions.

*Polymer-ion interaction*

FTIR spectral absorption modes of the host polymer (PEO) shows clear evidence of interaction of electron donor group of the polymer chain with cation. The shift in the position of COC stretching mode at 1048 and 1113 $cm^{-1}$ towards lower wavenumber side confirms the presence of strong interaction with salt (Table 3). The vibrational mode responsible for the mode at 1111 $cm^{-1}$ is due to stretching COC of host polymer (Fig 6). Besides, the absorption peaks at 1244 $cm^{-1}$ show an opposite trend with a shift towards the higher wavenumber side. Also, the wagging mode of host polymer at 1351 $cm^{-1}$ shifts toward lower wavenumber side on the addition of salt. Peak near 1280 $cm^{-1}$ corresponding to twisting asymmetric $CH_2$ mode disappears on complexation of a polymer blend with salt. Scissoring peak of polymer blend near 1460 $cm^{-1}$ in undisturbed on the addition of salt in a polymer matrix (Figure 7). The above analysis shows changes in intensity, shape and peak which strongly evidence the complex formation of a polymer blend with salt via polymer-ion interaction.

Figure 8(a–e) shows the spectral pattern of hexafluorophosphate ($PF_6^-$) group in the solid polymer electrolyte films observed in the wavenumber region 820-870 $cm^{-1}$. The investigation of peak profile of $PF_6^-$ a band with octahedral symmetry ($O_h$) of polymer salt system (Fig 8 a) suggests the presence of asymmetry in the characteristic band with maxima at 842 $cm^{-1}$ corresponding to $v_3(t_{1u})$ mode [32, 45-46]. This may be attributed to the distortion due to bending vibration of P-F bond produced by the ion-ion ($Li^+$ and $PF_6^-$) interactions, and it may be confirmed regarding the simultaneous presence of free ions/ion pairs by deconvoluting $PF_6^-$ vibrational mode. Any possible coordination of $PF_6^-$ group with cation usually results in lowering of the $PF_6^-$ group symmetry from $O_h$ point group to $C_{3v}$ or $C_{2v}$ depending on the nature of the coordination that $PF_6^-$ group possesses (tridentate/bidentate) with cations ($Li^+$) and other such active sites having Lewis acid character. However, the greater difference in charge between $Li^+$ and F atoms (0.3968) in the $C_{3v}$ structure suggests a stronger coordination than in $C_{2v}$ structure (0.3772). Nanofiller addition in the polymer salt matrix may lead to an interaction of fillers with the anions ($PF_6^-$) and changes are evidenced in terms of the shift in peak. Ion filler interaction is also investigated in next section for more detailed interaction in the various systems.

The free $FP_6^-$ anion has 15 vibration modes and can be represented as:

$$\Gamma = a_{1g} + e_g + 2t_{1u} + t_{2g} + t_{2u}$$

Of these modes, $v_1(a_{1g})$, $v_2(e_g)$ and $v_5(t_{2g})$ are Raman active, $v_3(t_{1u})$ and $v_4(t_{1u})$ are IR active, while the $t_{2u}$ is inactive where $v_1 = 741$, $v_2 = 567$, $v_3 = 838$, $v_4 = 558$ and $v_5 = 470$ $cm^{-1}$ [47].

*ion-ion interaction*

The effect of various nanofiller on the vibrational mode of $PF_6^-$ mode is visible in Fig 8 (b-e). The peak shows a change in shape, position and intensity with the addition of nanofiller. The plot shows no major change in the peak of $PF_6^-$ mode on the addition of nanofiller but for erbium oxide nanofiller shift in peak is clearly visible suggesting the interaction of the ion with the nanofiller due to large surface area. This causes a change in free anion area of the nanofiller based system as compared to the polymer salt system. This change in the free anion area, the peak position is attributed to the interaction between nanofiller and $PF_6^-$ anion evidenced by deconvoluting the hexafluorophosphate ($PF_6^-$) anion band confirming the presence of free ions and ion pairs.

The typical vibrational mode of $PF_6^-$ anion in SPEs films have been observed in wavenumber region ~830-850 cm$^{-1}$ and this shows asymmetry in peak as clearly visible in PS film (Figure 9 a). This peak asymmetry in $\upsilon_3(t_{1u})$ mode is the result of a loss in degeneracy from octahedral symmetry $(O_h)$ to $C_{3v}$ arising due to simultaneous presence of more than two components i.e. free ion, ion pairs. So to study above degeneracy and presence of both ions, deconvolution of $PF_6^-$ peak is done using Peak Fit (V 4.02) software using Voigt Amp profile. The best fit is said to be measured by the correlation coefficient ($r^2$). A quantitative estimation of fraction free ion and ion pair in the deconvoluted pattern is obtained using equation 5 & 6:

$$Fraction\ of\ free\ anion = \frac{Area\ of\ free\ ion\ peak}{Total\ peak\ area} \quad (5)$$

and

$$Fraction\ of\ ion\ pair = \frac{Area\ of\ ion\ pair\ peak}{Total\ peak\ area} \quad (6)$$

The deconvoluted pattern of SPE films with different filler concentration is shown in figure 9. The deconvoluted pattern appearing in PS film shows two distinct peaks one is at lower wavenumber and another at high wavenumber. It is fairly known that peak at lower wavenumber is attributed to free ions and at higher wavenumber is due to ion pairs [29]. In the case of polymer salt film without nanofiller (Figure 9; PPL) free ion peak is at 839 cm$^{-1}$ and ion pair at 842 cm$^{-1}$ which is the result of peak asymmetry. The deconvoluted pattern for SPEs shown in Figure 9 depicts the changing profile of free ion and ion pair peak in terms of change in position and area with a change in the type of nanofiller. A relative comparison of corresponding free ion area and ion pair area is summarized in Table 4. As listed there is relatively higher free ion area (57.07 %) for $Er_2O_3$ dispersed nanofiller as compared to polymer salt (50.58), $BaTiO_3$ (51.90 %), $CeO_2$ (54.45 %). This increase in area may be attributed to separation of more Li ion from $PF_6^-$ anion which leads to release of more free charge carriers and it shows the presence of enough interaction of nanofiller erbium oxide with salt supporting more dissociation of salt [48-49].

*Polymer- ion-nanofiller interaction*

For the pure PEO polymer broad band observed between 2840 cm$^{-1}$ and 2980 cm$^{-1}$ is attributed to symmetric/asymmetric stretching mode of $CH_2$. However, band splitting is seen as in figure 10 into two bands one near 2880 cm$^{-1}$ and another 2910 cm$^{-1}$ corresponding to symmetric $CH_2$ stretching ($\nu(CH_2)_s$), asymmetric $CH_2$ stretching ($\nu(CH_2)_a$) respectively. On addition of salt, both symmetric and asymmetric stretching mode alters their position indicating evidences the interaction of salt with polymer host. The deconvolution of the peak is done to get the original peaks in given wavenumber range by Voigt Amp fit. Fig 10 shows the presence of two peaks for stretching mode. Further, the addition of nanofiller shows a change in band position of stretching mode in terms of the shift in wavenumber. The increase in peak area for erbium oxide based SPE suggests the enhancement in conductivity provides by the stretching of the polymer chain. For titanium oxide based SPE reduction in peak area suggests the less stretching in the chain and hence shows a reduction in conductivity. So above study concludes that stretching of $CH_2$ group was interrupted by the addition of nanofiller in the polymer salt matrix.

This shift in band position shows the effect of nanofiller and presence of various interactions between polymer, salt and nanofiller in solid polymer electrolytes. Generally, these interactions impact strongly the dissociation of salt which results may improve free ions and results in faster ion dynamics in polymer nanocomposite films. This is further studied in the following section and is correlated with electrical properties.

**Complex impedance spectroscopy analysis**

Complex impedance spectroscopy is a useful technique in analysing the electrode electrolyte interface properties in determining equivalent circuit and circuit parameters. For pure polymer blend semicircle nature of curve is obtained with a diameter $R_b$ extending along the real axis from the origin as in Figure 11. The log-log presentation of these complex impedance plot is represented for better clarity and comparison of the different impedance plots in a single plot. It can be noticed that the semi-circular arcs are distorted but logarithmic plots are superior in various aspects [50]. Figure 11 shows typical complex impedance spectra (log-log - plot) of salt containing polymer blend/PNC films sandwiched between two stainless steel electrodes corresponding to the bulk impedance of sample. It is well known that high frequency response tells information about properties of an electrolyte such as bulk resistance and the low frequency response carries information about the electrode/electrolyte interface. The arc at low frequency side represents double layer capacitance due to the migration of ions at low frequency at electrode/electrolyte interface and the disappearance of high-frequency semicircle reveals that total conductivity is mainly due to the lithium ions. The double layer capacitive effect is due to the non-faradic process occurring at the interface of SPE films and stainless steel electrode. It also indicates the inhomogeneous nature of the electrode–electrolyte interface

The identical trend is obtained for all nanofiller dispersed polymer electrolyte films. In ideal situations, the impedance plot in low-frequency regions is expected to show straight line parallel to the imaginary axis. However, the double-layer capacitive effect resulting as result of the blocking electrodes results in a semi-circular arc. The intersection of impedance spectra with the real axis represents bulk electrolyte resistance ($R_b$) for transportation of Li ions. At low frequencies, there is sufficient time to accumulate the space charge across the electrode electrolyte interface that results in an enormous increase in the measured capacitance [51-53]. This type of behaviour proposes that migration of ions occurs via free volume available in solid polymer electrolytes and is represented by a resistor. The low-frequency response of CIS pattern remains almost identical for the SPE films irrespective of different filler and results from the accumulation of space charge carriers at the interface of the ionically conducting SPE films and stainless steel blocking electrode.

An electrical equivalent circuit shows a parallel combination of constant phase element (CPE) and a resistance connected in series with another constant phase element (CPE). Former one represents the bulk properties of solid polymer electrolyte sample and later one double layer between electrode/electrolyte interfaces (as shown in inset).

The presence of a CPE in the material sample reveals the multiphase character comprising of a microstructure having both crystalline, amorphous and a mixture of the two phases in the PS complex film [17, 54]. Constant phase element indicates the value of capacitance and $n_1$ represents the deviation from ideal semi-circular behaviour and $n_2$ is the slope of low frequency spike [55]. From the fitting pattern of CIS (Figure 11) it is observed that the fitted (solid line) and experimental pattern is in good agreement with each other that validate the authenticity of the investigation. Change in fitting parameters with various nanofiller shows a change in electrical properties due to their different particle size and Scherer length as evidence from the XRD. The parameters given in Table 4 provide evidence for a change in the sample electrical response with change in nanofiller in the SPE films. The model of the electrical equivalent circuit remains identical in a pattern for the SPE and PS films with essential difference in the values of bulk resistance ($R_b$), constant phase elements $Q_1$ and $Q_2$ and their exponent's $n_1$ and $n_2$, respectively as recorded in Table 5.

*Electrical conductivity*

The ionic conductivity is one of the most important performance measures for SPEs and furthermore, a proper analysis of ionic conductivity data can reveal details about the conduction mechanism. The electrical conductivity of the SPEs films has been calculated from the complex impedance spectrum given in equation 2. The intercept of the high frequency semi-circular arc or spike with the real axis gives an estimate of the bulk d. c. resistance ($R_b$) [56]. The ionic conductivity of pure polymer blend was $6.4\times10^{-10}$ S cm$^{-1}$. The electrical conductivity increases with the addition of salt due to the availability of ions for migration. Further, the addition of nanofiller enhances the salt dissociation and continues increase is seen. The highest ionic conductivity was $2.3\times10^{-5}$ S cm$^{-1}$ for erbium oxide nanofiller and almost five order greater than pure polymer blend and corresponds to the complete dissociation of salt in evidenced by the FTIR spectra. Besides with addition of $TiO_2$ nanofiller conductivity decreases which is consistent with the reduction of dissociated free ions calculated by deconvolution of FTIR spectra and also the agglomeration of particles restrict the ion migration as noticed in the in FESEM micrograph [57]. The typical conductivity variation with different nanofiller agrees well with the filler dependent changes in the fraction of the free anion in the PSPEs system.

Although the nanofiller content was kept constant (10 wt. %) in this study but from their different particle size and dielectric constant value it is expected that it may affect the polymer salt matrix differently. Although the dielectric constant is higher (~1700) for the $BaTiO_3$ the large contrast in the dielectric constant of the polymer and nanofiller results in the inhomogeneity of the polymer nanocomposite matrix and that does not enhance the conductivity so much which shows a negative effect. The dielectric constant of $Er_2O_3$ is lowest (~12) among all but has highest ionic conductive that may be due to homogeneity in the electric field of both polymer and nanofiller owing to the small difference in the dielectric constant [32, 58]. The lower dielectric constant value supports the complete dispersion of nanofiller properly with the polymer salt matrix and can play its effective role in the enhancement of the ionic conductivity while the high dielectric constant nanofiller is unable to show its effectiveness in enhancing the conductivity and may be due to agglomeration of nanofiller. So, the overall high dielectric constant of the polymer nanocomposite is desirable for the fast ion transport [59]. Further insight into understanding the overall dielectric constant will be provided by the dielectric spectroscopy study which will be explored in detail in next section.

An excellent correlation of maxima in the fraction of free ion variation and conductivity with nanofiller content is in support of our scheme (Figure 12 (a, b)). The ionic conductivity of an electrolyte is related to the number of the charge carriers (n), ionic charge (q), and ion mobility (µ) in the electrolyte expressed as follows (eqn 7):

$$\sigma = nq\mu \qquad (7)$$

Where n is a fraction of free ion involved in ionic transport, q is ionic charge and µ (ion mobility) is strongly influenced by polymer chain segmental mobility. As q and µ are constant for a particular matrix so conductivity only depends on the number of free charge carriers as studied in followed section [60]. Further, to support the present results various important transport parameters are calculated and correlated with ionic conductivity as in the following section.

*Correlation of free ions area (%), ionic conductivity, number density (n), mobility ($\mu$), diffusion coefficient (D) and viscosity ($\eta$)*

The desirable property of a solid polymer electrolyte is high ionic conductivity (σ) which is directly linked with number density (n), mobility ($\mu$) of charge carriers and diffusion coefficient (D) for any plastic separator stands for electrolyte cum separator. Transport of lithium ions gives rise to conduction on the application of external potential. As from conductivity results it is observed that conductivity increases with addition of salt in pure polymer blend and further addition of nanofiller leads to enhancement with maxima for erbium oxide nanofiller. This enhancement in conductivity may be attributed to the increase of mobility or number density of charge carriers [61]. The mobility (($\mu$) of ions represents the degree of ease with which ions pass through media when an external electrical field is applied, and the diffusivity (D) represents the ease with which ions pass through media under a concentration gradient [62]. Both parameters depend on the number of free charges as ion pair formation or association may reduce the above number and mobility. As diffusivity is quite tedious to calculate experimentally, so impedance study directly provides this value using ionic conductivity. So it becomes important to study the number density of mobile ions and their mobility for performance and development of a good solid polymer electrolyte. Here the above parameters are calculated using the FTIR spectroscopy as proposed by Arof et al., [63].

The variation in conductivity can be related to the number density (n), mobility ($\mu$) and diffusion coefficient (D) of charge carriers in the electrolyte. Diffusion is said to occur when ion jumps from one site to neighbouring coordinating site by exchange of position. FTIR deconvolution was done to determine the percentage area of free ion and ion pair and the areas are plotted as a function of various nanofiller (Figure 12). The area of free ions increases with addition of nanofiller and is highest for erbium oxide nanofiller. The number density (n), mobility ($\mu$), diffusion coefficient (D) and viscosity ($\eta$) of the mobile ions were calculated using following eqn (8 a-d):

$$\begin{cases} n = \dfrac{M \times N_A}{V_{Total}} \times free\ ion\ area\ (\%) & (8\ a) \\ \mu = \dfrac{\sigma}{ne} & (8\ b) \\ D = \dfrac{\mu k_B T}{e} & (8\ c) \\ \eta = \dfrac{k_B T}{6\pi r D} & (8\ d) \end{cases}$$

In eqn (8 a), M is the number of moles of salt used in each electrolyte, $N_A$ is Avogadro's number ($6.02 \times 10^{23}$ mol$^{-1}$), $V_{Total}$ is the total volume of the solid polymer electrolyte, and $\sigma$ is dc conductivity. In eqn (8 b), e is the electric charge ($1.602 \times 10^{-19}$ C), $k_B$ is the Boltzmann constant ($1.38 \times 10^{-23}$ J K$^{-1}$) and T is the absolute temperature in eqn (8 c). $\eta$ is viscosity in eqn (8 d) and r is radius of diffusing species (0.6 Å for Li). Table 6 lists the values of $V_{Total}$, free ions (%), n, $\mu$, D obtained using the FTIR method [63]. It is observed that the ionic conductivity is closely related to the number density of mobile charge carriers. Figure 12 reveals the one-to-one correspondence between free ion area (%), ionic conductivity ($\sigma$), charge carrier number density (n), charge carrier mobility ($\mu$), diffusion coefficient (D) and viscosity ($\eta$). The trend in variation in free ions area (%), n, $\mu$ and D is almost similar. It can be concluded that erbium oxide doped polymer electrolyte increased the dissociation of salt resulting in an increased number density of mobile ions, formed more amorphous phases leading to an increased ionic mobility and the diffusion coefficient of the mobile ions. The reduction in viscosity further supports our achievement of highest conductivity and number of free ions for erbium oxide nano filler based solid polymer electrolyte [64]. Above results are strongly in favour of desirable SPEs for energy storage devices with the evidence provided by the FTIR and impedance analysis. Further to get the understanding of the dielectric constant and the particle size of the nanofiller dielectric analysis is performed in the following section.

**Dielectric Spectroscopy Analysis**

Dielectric analysis of solid polymer electrolyte materials is desirable to investigate in detail the transport of ion and are explained in terms of the real and imaginary parts of complex permittivity ($\varepsilon^*$). The complex dielectric permittivity (real part of dielectric permittivity $\varepsilon'$ and imaginary part of dielectric permittivity $\varepsilon''$) is plotted against the frequency in the range 1 Hz to 1 MHz for different nanofiller. The dielectric permittivity describes the polarizing ability of a material in the presence of an applied external electric field. As permittivity is a function of frequency so it is a complex quantity shown below (eq 9):

$$\varepsilon^* = \varepsilon' - j\varepsilon'' \quad (9)$$

The real part of dielectric permittivity ($\varepsilon'$) is proportional to the capacitance and measures the alignment of dipoles, whereas the imaginary part of dielectric permittivity ($\varepsilon''$) is proportional to conductance and represents the energy required to align the dipoles. Here $\varepsilon'$ is related to the stored energy within the medium and $\varepsilon''$ to the dielectric energy loss of energy within the medium. The real and imaginary parts of the dielectric permittivity are evaluated using the impedance data by eqn 10 a & b :

$$\begin{cases} \varepsilon' = \dfrac{-Z''}{\omega C_o (Z'^2 + Z''^2)} & (10\ a) \\ \varepsilon'' = \dfrac{Z'}{\omega C_o (Z'^2 + Z''^2)} & (10\ b) \end{cases}$$

where $C_o$ ($=\varepsilon_r A/t$) is the vacuum capacitance, $\varepsilon_r$ is the permittivity of free space and $\omega$ is the angular frequency. The complex dielectric permittivity (dielectric constant $\varepsilon'$ and dielectric loss $\varepsilon''$) as a function of frequency with different nanofiller in log-log scale are shown in Figure 13. The value of $\varepsilon'$ & $\varepsilon''$ at low frequencies is three to four time higher than the value at high frequency. This increase in low frequency window indicates the dominance of electrode polarization (EP) effect due to the accumulation of the long path travelled charge carriers near electrodes [65-66]. This results into the three to four order increase in the dielectric permittivity value as compared to the pure polymer blend. Further, both the $\varepsilon'$ & $\varepsilon''$ value shows dispersion with an increase of frequency above 1 kHz which may be attributed to the bulk dielectric properties of the investigated materials (low frequency window). A close look at Figure 13 a indicates that the all the dielectric spectra for pure polymer and polymer salt with different nanofiller are following the identical trend. But, the value of dielectric constant increases with the addition of nanofiller as compared to polymer salt complex and may be attributed to the release of more number of charge carriers as evidenced by the FTIR analysis. As after the dispersion of the nanofiller of different particle size and dielectric constant various possible interactions influence the dielectric constant. The highest dielectric constant or the maximum dielectric polarization was for the polymer salt complex dispersed with the $Er_2O_3$ nanofiller and follows the trend, $Er_2O_3$> $TiO_2$> $BaTiO_3$> $CeO_2$> no filler (polymer salt)> pure polymer blend. The polymer-ion-nanofiller interaction as evidenced by the FTIR weakens the polymer-ion and ion-ion interactions. This alteration enhances the polarization ability of the whole polymer matrix which leads to high dielectric constant. Now, in the high frequency window both $\varepsilon'$ & $\varepsilon''$ exhibits the dispersion and is attributed to the bulk dielectric properties of the present system.

The dielectric analysis was explored further in terms of the particle size and dielectric permittivity. In the present investigation, although the nanofiller content was fixed at 10 wt. % but they possess different particle size and dielectric constant. As the dielectric constant of the whole polymer matrix was three to four orders higher than the pure polymer salt system at a low frequency that indicates that the nanofiller strongly affects the polymer chain arrangement. The dielectric constant values are plotted against the crystalline size and the $\varepsilon_r$. From the figure 14 a it is observed that the both L and $\varepsilon_r$ shows an atypical trend with the dielectric constant. The dielectric constant was higher for the $Er_2O_3$ nanofiller while its crystalline size is almost half than of $BaTiO_3$ which has a high dielectric constant (Figure 14 b). It can be concluded that beside the particle size and the ion-nanofiller interaction some other factor also modifies the polymer chain arrangement and hence the interaction. Figure 14 b shows that there is a decrease of the dielectric constant $\varepsilon'$ for the $BaTiO_3$ and that may be associated with the large contrast in the dielectric constant of the polymer and nanofiller which leads to inhomogeneity in the polymer matrix. The dielectric constant of whole polymer matrix was also plotted against the particle size (Figure 14 c). It may be noticed that, although the particle size of the $Er_2O_3$ and $BaTiO_3$ are almost same, both effects the polymer matrix in a different way.

The low frequency response is coined by electrode polarization (EP) effect which is due to the formation of electric double layer capacitances due to free charge build up at the electrolyte/electrode interface. While the higher frequency region prevents the dipole structures to get sufficient time for fast response in changing ac electric field and hence there is a decrease in a number of dipoles which contribute to EP. This prevents the diffusion of ions in the direction of the field and results in a decrease in dielectric constant ($\varepsilon'$). Figure 13 b shows the large value of dielectric loss ($\varepsilon''$) towards low frequency region and may be attributed to the accumulation of free charge carrier at electrode/electrolyte interface because in this region charge carriers get sufficient time to accumulate at electrode/electrolyte interface and contribute to large dielectric loss [67]. The overall dielectric constant is higher for the polymer salt complex dispersed with the erbium oxide nanofiller.

## Modulus Spectroscopy analysis

Figure 15 a & b shows the electric modulus (real part $M'$ and loss $M"$) spectra of the PEO-PVC blend-based electrolyte films prepared through solution casting methods. These spectra are commonly plotted and analyzed for the solid dielectric materials, as they are free from the contribution of electrode polarization effect at low frequencies, and also independent of the type of electrodes material, and the adsorbed impurities in the sample [68].

In case of ionic conductors, at low frequency region EP is reflected by a significant increase in the dielectric loss i.e. $\varepsilon"$ spectrum. At low frequency side of spectra low value of $M'$ is observed and a non-linear behavior is seen with increase in the frequency. At low frequencies the observed values of dielectric permittivity do not refer to the bulk of the material; this is related to the so-called "interfacial polarization" dominant in these composites polymeric materials. To avoid contributing of the interfacial polarization, the modulus formalism can be used for analyses of the dielectric behavior of solid polymer electrolytes and is represented in terms of complex modulus ($M^*$) which is an inverse of the complex permittivity ($\varepsilon^*$) calculated using the formula shown below (equation 11 a & b):

$$\begin{cases} M^* = M' + i\,M" & (11\,a) \\ M^* = \dfrac{1}{\varepsilon^*} = \dfrac{\epsilon'}{\varepsilon'^2 + \varepsilon"^2} + i\dfrac{\epsilon"}{\varepsilon'^2 + \varepsilon"^2} & (11\,b) \end{cases}$$

Where $M'$ and $M"$ are the real and imaginary parts of the dielectric modulus.

At low frequency imaginary part of modulus, spectra show zero value which indicates the negligible electrode polarization due to lack of restoring forces leading the mobility of charge carriers under the action of an induced electric field and supports the long-range mobility of charge carriers [69-70]. Later with an increase in frequency value of $M"$ increases and maximum at high frequency contributed by bulk properties of SPEs. A peak is observed in pure polymer blend sample and peak shifts to high frequency with the addition of salt (Fig 15 b). After addition of nanofiller clear peak is not visible but variation in graph conforms the presence of peak at high frequency side as compared to salt doped polymer electrolyte (Fig 15 b). The presence of long tail in all SPEs at low frequencies is attributed to their large capacitance values associated with the electrodes. Shift of peak towards high frequency side suggests decrease in relaxation time ($\tau_0$) which directly supports our idea of enhancement in conductivity and mobility of charge carriers.

## ac conductivity analysis

The AC electrical measurements (AC conductivity) of PEO/PVC based solid polymer electrolyte were obtained at a frequency range from 1 Hz to 1 MHz at room temperature. The frequency variation of real part of conductivity is shown in Figure 16. A common feature in polymeric electrolytes shows that ac conductivity increases with increasing the applied frequency due to fast ion migration. It may also attribute to increase the absorbed energy which leads to increase in a number of the charge carriers leads to charge build up. The frequency dependent real part of conductivity shows three distinct regions, firstly low frequency dispersion region then frequency independent plateau region and followed by a high frequency dispersive region. The low-conductivity value at the low frequency dispersion region is related to the accumulation of ions (electrode polarisation) due to the slow periodic reversal of the electric field and commonly occurs in solid polymer electrolyte materials. This region disappears with an increase in frequency. The intermediate region at slightly higher frequency corresponds to the frequency independent plateau region from which the room temperature dc conductivity can be obtained which is the result of long range diffusion of ions. At last the high frequency dispersion region occurs corresponding to a bulk relaxation phenomenon due to short range ion transport associated with ac conductivity. Pure polymer and polymer salt system shows all the three regions. All system shows a shift in both intermediate frequency region and high frequency region towards high frequency. For the high conductive system, high frequency dispersion region corresponding to bulk relaxation phenomena falls outside the measured frequency range and hence could not be observed. The dc conduction is attributed to random hopping of ions between localized states, while the cause of AC conduction is correlated ion hopping in high frequencies region. This observation is indicative of complex ion transport process possibly due to the combined effect of space charge polarization (electrode polarization at the electrode electrolyte interface) at low frequency followed by long-range ion migration at high frequency [60, 66]. AC conductivity is calculated by equation 12:

$$\sigma_{ac} = \omega \varepsilon_o \varepsilon'' = \omega \varepsilon_o \varepsilon' \tan \delta \quad (12)$$

Where $\omega$ is the angular frequency, $\varepsilon_o$ is the dielectric permittivity of the free space and $\varepsilon''$ represents the dielectric loss. The conductivity for pure polymer blend is very low as observed in figure 16 (a) and is due to EP effect causing a decrease in a number of free charge carriers. The addition of salt shifts the curve toward high frequency region which is evidenced by an increase in conductivity. Also, the different slope for all system shows varying strength of interface polarisation effect. The electrode polarization contribution in conductivity is clearly observed by the strong dispersive effect which is typical to a fast ion conductor and saturation limit of electrode polarization effect, evidence change in the mechanism of ion migration. This particular corresponding threshold frequency of transition shows the dc limit or long-range migration. Two types of phenomena long-range conduction due to the migration of free cations and anions in polymer nanocomposites and preferred site hopping conduction through the polymer hetero sites are responsible for ac conductivity in solid polymer electrolytes [71]. The high frequency region is visible in pure polymer blend and polymer salt system while for another sample it falls outside the measured frequency range. The high frequency region in both systems follows well known Jonscher's power law (JPL) given by en. 13

$$\sigma' = \sigma_{dc} + A\omega^n \quad (13)$$

$\sigma'$ and $\sigma_{dc}$ are the AC and DC conductivities of electrolyte, while A and n are the frequency independent Arrhenius constant and the power law exponent, respectively. Basically, ion migration at low frequency is faster and jump from one available site (ether group) to another in the host polymer matrix when the frequency is lower than the hopping frequency ($\omega_p$). The low frequency results in long relaxation time, and ion contribution is seen in dc conductivity. Two competing hopping processes are observed at high frequency: (i) jumping of ion back to its initial position (correlated forward–backward–forward), i.e., unsuccessful hopping and (ii) the neighbourhood ions become relaxed with respect to the ion's position (the ions stay in the new site), i.e., successful hopping [72]. The increase in number of successful hopping ($\omega > \omega_p$) results in a more dispersive ac conductivity at higher frequency region [73]. When frequency exceeds $\omega_p$, $\sigma'$ increases proportionally on, where n < 1. Low chi-squared value of order of $10^{-13}$ suggests best JPL fit for the experimental results and fitted values are shown in inset (Figures 16 b & c). The frequency-dependent region fitting provides an estimate of the pre-exponential factor (A) and fractional exponent (n). Value of n between 0.5 and 1 suggests that SPE system is a pure ionic conductor. The power law behaviour is universal property of materials. The $\sigma_{dc}$ (dc conductivity) value obtained from JPL fitting for pure polymer and polymer salt system is $2.08\times10^{-10}$ S cm$^{-1}$, $4.66\times10^{-7}$ S cm$^{-1}$ and is in good agreement with value obtained using the bulk resistance ($R_b$) from impedance plot given as; $6.40\times10^{-10}$ S cm$^{-1}$ and $7.10\times10^{-7}$ S cm$^{-1}$ respectively.

**Capacitive behaviour analysis**

The real and imaginary part of capacitance is calculated using the following equation 14 a-b.

$$C' \begin{cases} C'(w) = -\dfrac{Z''(w)}{w|Z(w)|^2} & (14\,a) \\ C''(w) = \dfrac{Z'(w)}{w|Z(w)|^2} & (14\,b) \end{cases}$$

Where, Z′ and Z″ are the real and imaginary part of the impedance, ω is the frequency. Here, C′ corresponds to the static capacitance of the system corresponding to electrode/electrolyte interface when measured at low frequency alternating current and C″ corresponds to dielectric losses in the form of energy dispersion of the electrolyte. The variation of real and imaginary parts of the capacitance with the frequency for different nanofiller is shown in Fig. 17. For systems under study, the value of capacitance at lower frequency changes with salt concentration and is highest for best conducting sample, i.e. With the increase in frequency, C′ sharply decreases with a saturation region above 10 kHz. The high value of capacitance at low frequency may be attributed to the contribution of more free ion in the system while at high frequency only the ions at electrode/electrolyte surface contribute [73]. The sharp peak in the graph of the frequency dependent imaginary C″ corresponds to the energy dissipation and plot of imaginary C vs frequency is shown in Fig. 17. C″ passes through a maximum at a particular frequency ($f_o$) which corresponds to the minimum time ($\tau_o$) required to discharge all the energy from the device with an efficiency >50%, called for the dielectric relaxation time. The presence of a peak at low frequency is seen in Fig 17. While in other samples peak can be achieved at the further low frequency.

**Ionic transference number**

For SPEs to be used in rechargeable battery applications, the principal charge carriers should be ions, and the fractional contribution of ionic conductivity to the total conductivity should be as close to unity as possible while the fractional contribution of the electronic conductivity should be negligibly small (close to zero). Therefore, it is necessary to determine the transference number due to ions and electrons. The transference number of an ion is the fraction of the total current that is carried by the respective ion across a given medium. The total effective ionic transference number of SPEs has been studied by separation of ion and electronic contribution. Variation of current as a function of time for the PS film and SPE films with different filler at room temperature are recorded as shown in Figure 18.

The current decays immediately and asymptotically approach steady state. The pattern shows two current one is initial current known as total current ($i_t$) and after that there is a sharp drop in the value of current with the passage of time. After some time current get a saturated and same trend of constant current is obtained for all the samples. The two processes of migration of ions under the influence of an external field and diffusion due to concentration gradients are hostile, and therefore after sufficiently long time, an establishment of steady state is reached. Initially in SPE membrane high current is attributed to migration of both electrons and ions while current after cell polarization due to blocking electrodes (SS) is attributed to movement of electrons only, known as residual electronic current ($i_e$). The initial decrease in current until the saturation state is primarily due to the formation of the passivation layer on electrodes. Now, a concentration gradient of ions develops and opposes the applied current by diffusion of ions. As the polarization builds up because of the applied voltage, the ions were blocked at the blocking electrode thereby blocking the ionic current ($i_{ion}$), and the final current comprises only the electronic current ($i_e$). A large transference number also indicates a reduction in concentration polarization [49]. The estimated value of the ionic transport number ($t_{ion}$) using experimental data of $i_t$ and $i_e$ by the eq. (3 & 4) and is given in Table 7. Further ionic and electronic conductivity contribution is calculated using transport number using equation 15-17.

$$i_t = i_{ion} + i_{elec} \quad (15)$$

The ionic transference number has also been used to estimate ionic and electronic contributions to the electrical conductivity.

Therefore

$$\sigma_{ionic} = \sigma_{electrical} \times t_{ion} \quad (16)$$

$$\sigma_{electronic} = \sigma_{electrical} \times t_{electronic} \quad (17)$$

The transport number values clearly indicate that the SPE samples are a chiefly ionic conductor with $t_{ion} \approx$ 0.95 negligible contribution from electron and may be sufficient to fulfil the requirement of solid state electrochemical cells. The ionic conductivity and electronic conductivity are also in correlation with electrical conductivity value.

**Voltage stability**

The working voltage or electrochemically stability window of the SPEs membranes has been studied by observing the variation of current and voltage (I & V) using LSV technique (Figure 19) for PS films and SPE films with nanofiller. The irreversible onset of the current determines the electrolyte breakdown voltage, which reveals the electrochemical stability of the system where no oxidation or reduction takes place. It indicates a gradual increase in the magnitude of current with the rise in the applied voltage across the cells. Initially there is almost constant current through the electrodes, while after that with an increase of voltage applied across cell configuration there is a slow increase in current and at some voltage beyond some limit abrupt change occurs in current which gives the voltage window of the membrane. The voltage value is obtained by the extrapolating point of intersection of two extrapolated lines of the high-voltage linear current trace and the low-voltage zero current trace parallel to the X-axis. PS complex sample seems to influence the oxidation kinetics in the cell. The addition of filler in PS films gives a safe operating range with the majority of the lithium battery electrode couples. During the LSV, the working electrode is polarized and the onset of the current may be taken as the decomposition voltage of the given polymer electrolyte. Further, the absence of negative current also indicates that there was no decomposition on the working stainless steel electrode [74-75]. A comparison of this with PS film and indicates an enhancement in voltage window in comparison to $TiO_2$, $CeO_2$, $BaTiO_3$ with a maximum value of ~3.5 V for $Er_2O_3$ nanofiller. The broadness in stability window may be due to the addition of filler in PS films. As filler attachment with PEO chain occurs and makes SPE film to have high decomposition potential.

In addition, the electrochemical stability window was investigated by the cyclic voltammetry analysis also between voltage range -2.5 V to + 2.5 V. To achieve high energy density (E = $CV^2/2$), the high value of capacity (decided by electrode material) and working voltage (depends upon stability window of the electrolyte) are required. Since E depends upon the square of the voltage, hence it becomes an essential and desirable parameter for material development and application purpose [76]. The electrochemical stability window has been found to be approx. 4 V for all the samples and is highest for erbium oxide doped polymer nanocomposite (Figure 20 e).

**Scheme**

To understand the individual role of the polymer, salt and the nanofiller a scheme is proposed (Figure 21) that will enable the clear understanding of the enhanced electrical and dielectrical parameters. The polymer host used in the present investigation is PEO and it consists of an electron donor rich group (ether group). When the salt is dissolved in the polymer blend then the salt get dissociated in the cation and anion. Further dissociated cation seems to be coordinated with ether group and anion with methylene group of the of host polymer chain. This coulombic interaction separated the ion pairs in $Li^+$----$PF_6^-$. Since, the cation is Lewis acid so it has strong tendency to attach with the Lewis base which in the present scenario is the ether group of the polymer chain. This cation gets coordinated with the ether group while anion gets attached with methylene group in the polymer backbone. Here, methylene having H group acts as Lewis acid and get coordinated with the anion via hydrogen bonding. Now, the creation of new coordinating sites and breaking of previous takes place that leads to migration of the cation via the polymer chain. Also, the coordinating interaction of the cation with polymer chain modifies the polymer chain arrangement and disorder is produced that evidences the increase in the polymer chain flexibility. The enhanced flexibility is an indication of the enhanced conductivity and the fast segmental motion of polymer chain provide a path for ion transport.

Now, the dispersion of nanofiller in the polymer salt matrix is done and various interaction such as polymer-ion and ion-ion are influenced while some new interaction comes into picture such as polymer-ion-nanofiller, polymer-nanofiller. The presence of the above interaction changes the previous interaction occurring in the polymer salt complex. As, nanosize fillers are used in the present system that directly leads to an enhanced surface area of the interaction. Also, the nanofiller is having the surface group on the outer surface that welcomes the new inclusion of new interactions. The negative charge is due to the oxygen of the surface group and this is beneficial for the faster ion migration. As oxygen is a Lewis base so cation has the possibility to interact with this alongwith the ether group of the oxygen. There may be competition between the surface group and the ether group for interacting with the cation. This makes the polymer matrix unstable and the more disturbance in the arrangement of the polymer chain enhance the free volume required for on migration or it can be said that amorphous phase gets improved that is necessary for a fast ionic conductor. Since, the coordinating sites are crucial for the ion migration so now cation get additional sites which make the swim of cation at a faster rate as earlier. The nanofiller addition provides a conductive path to the cation that supports the smooth transport and takes part in conduction. As, all nanofiller have a different particle size and the dielectric constant so they interact in a different manner with the polymer salt system. But, the overall contribution of all nanofiller is to modify the structure of the polymer chain and provide additional coordinating sites so that cation may easily jump from one to another coordinating site.

**Conclusions**

Free standing polymer nanocomposite films comprising of (PEO−PVC)+LiPF$_6$ with 10 wt. % nanofiller (BaTiO$_3$, CeO$_2$, Er$_2$O$_3$ & TiO$_2$) have been prepared through solution cast technique. X-Ray diffraction results confirm the polymer nanocomposite formation. The FTIR study indicated clear evidence for polymer–ion, ion–ion, and polymer–ion–nanofiller interaction. The deconvoluted pattern of anion peak (PF$_6^-$) evidences maximum number of free ion for Er$_2$O$_3$ nanofiller. Impedance spectroscopy results provide evidence of bulk conduction as the major electrical transport process and the highest conductivity is ~2.3 × 10$^{-5}$ S cm$^{-1}$ for 10 wt. % Er$_2$O$_3$. Ion transport number (~0.99) results indicate the SPE films to be predominantly ionic with broad voltage stability window (~ 3.5 V). The number density (n), mobility ($\mu$), diffusion coefficient (D) and viscosity ($\eta$) of the nanofiller doped samples were found to increase with nanofiller up to erbium oxide and beyond this for titanium oxide doped, values of these three parameters decreased. The dielectric spectroscopy provides crucial information of increase in three to four order of the dielectric constant as compared to nanofiller free polymer salt matrix and decrease with increase of frequency with highest at low frequency window due to electrode polarization effect. Finally, a scheme is proposed to supports the experimental evidences and facilitate easy understanding of the role of nanofiller.

**Acknowledgement**

One of the authors acknowledges CUPB for financial support and partial financial support from UGC Startup Grant (GP-41). The author is also thankful to Mr. Dinesh Kumar research scholar in School of Materials Science & Technology at Indian Institute of Technology (BHU), Varanasi for support in XRD characterization.

**Figure Caption**

**Figure 1.** Molecular structures of the constituents.

**Figure 2.** Flow chart of solution cast technique

**Figure 3.** XRD pattern of (a) PEO-PVC main peaks of (PEO-PVC)-LiPF$_6$ + 10 wt.% nanofiller based SPEs films in range 2θ =10° and 30° and (b) major nanofiller peaks in range 2θ =30° and 50°.

**Figure 4.** FESEM micrographs of (a) PP, (b) PPL, (c) PPLB, (d) PPLC, (e) PPLE, (f) PPLT

**Figure 5.** EDX elemental analysis pattern of (a) PP, (b) PPL, (c) PPLB, (d) PPLC, (e) PPLE, (f) PPLT electrolyte membranes.

**Figure 6.** FTIR spectrum of SPE films comprising of (a) PP, (b) PPL, (c) PPLB, (d) PPLC, (e) PPLE, (f) PPLT

**Figure 7.** Evidences of polymer-ion-nanofiller interaction by changes in profile of wagging mode of CH$_2$ based on **(a)** PP, (b) PPL, **(c)** PPLB, **(d)** PPLC, **(e)** PPLE, and **(f)** PPLT.

**Figure 8.** Effect of nanofiller concentration on the changes in the absorption band profile of PF$_6^-$ vibrational mode in (a) PPL, (b) PPLB, (c) PPLC, (d) PPLE, and (e) PPLT.

**Figure 9.** Typical deconvoluted pattern of hexaflorophospahte (PF$_6^-$) band at ϑ=840 cm$^{-1}$ showing contribution of free ion and ion pair in (a) PPL, **(b)** PPLB, **(c)** PPLC and **(d)** PPLE.

**Figure 10.** Evidence of polymer-ion-nanofiller interaction by change in spectral band profile of symmetric/asymmetric mode of CH$_2$ based on **(a)** PP, (b) PPL, **(c)** PPLB, **(d)** PPLC, **(e)** PPLE, **(f)** PPLT.

**Figure 11.** Complex impedance plot of SPE films **(a)** PP (b) PPL **(c)** PPLB **(d)** PPLC **(e)** PPLE **(f)** PPLT.

**Figure 12.** Plot of variation in free ions area (%), ionic conductivity (σ), charge carrier number density (n), charge carrier mobility ($\mu$), diffusion coefficient (D) and viscosity (η) obtained from FTIR method.

**Figure 13.** Variation of real (ε′) and imaginary part (ε″) of dielectric constant with frequency for (a) PP, (b) PPL, (c) PPLB, (d) PPLC, (e) PPLE, (f) PPLT.

**Figure 14.** Variation of ε′ (at 10 kHz) against the crystalline size, dielectric permittivity and the particle size of the various nanofiller.

**Figure 15.** Frequency dependent real part $M'$ and loss $M''$ of complex electric modulus for (a) PP, (b) PPL, (c) PPLB, (d) PPLC, (e) PPLE, (f) PPLT.

**Figure 16.** Variation of A.C. conductivity as a function of frequency for (a) PP, (b) PPL, (c) PPLB, (d) PPLC, (e) PPLE and (f) PPLT.

**Figure 17.** Variation of C′ and C″ with frequency for (a) PP, (b) PPL, (c) PPLB, (d) PPLC, (e) PPLE, (f) PPLT.

**Figure 18.** Ion transference number of SPE films (a) PPLB, (b) PPLC, (c) PPLE, and (d) PPLT with PPL in inset.

**Figure 19.** Linear Sweep Voltammetry of SPE films (a) PPLB, (b) PPLC, (c) PPLE and (d) PPLT with PPL in inset.

**Figure 20.** Cyclic voltamogram of (a)PP, (b) PPLB, (c) PPLC, (d) PPLE and (e) PPLT.

**Figure 21.** Proposed interaction scheme in the polymer nanocomposite matrix.

**Table Caption**

**Table 1**. Particle size and crystalline size of the $BaTiO_3$. $CeO_2$, $Er_2O_3$ and $TiO_2$ nanofiller used in the present system for preparation of polymer nanocomposites.

**Table 2**. Values of 2θ (degree), d-spacing (nm) and R (Å) of pure PEO-PVC and polymer salt with $BaTiO_3$, $CeO_2$, $Er_2O_3$ and $TiO_2$ for (120) diffraction peak.

**Table 3.** FTIR band identification and assignment in SPE films (a) PP (b) PPL (c) PPLB (d) PPLC (e) PPLE (f) PPLT

**Table 4.** Peak position of deconvoluted free ion and ion pair peak of SPE film

**Table 5.** Fitted Parameters of Nonlinear Least Squares (NLS) Fit of the samples comprising of SPE films

**Table 6**. The values of $V_{Total}$, free ions (%), n, $\mu$, D, $\eta$ obtained using the FTIR method.

**Table 7.** Ionic Transference Number, Electronic and Ionic conductivity for various nanofiller based solid polymer electrolytes